\documentclass[twocolumn,times,dvipsnames]{aastex631}

\usepackage[T1]{fontenc}
\usepackage{amsmath}
\usepackage{textcomp}
\usepackage{gensymb}
\usepackage{graphicx}
\usepackage{url}
\usepackage[caption=false]{subfig}
\usepackage{float} 
\usepackage{soul}
\usepackage{multirow}
\usepackage{newtxmath}

\newcommand{\noun}[1]{\textsc{\MakeLowercase{#1}}}

\newcommand{\code}{\texttt}

\newcommand{\m}[1]{$M_#1$}
\newcommand{\msun}{M$_\odot$}

\newcommand{\bafe}{[Ba/Fe]}

\newcommand{\yfe}{[Y/Fe]}

\hypersetup{colorlinks=True,linkcolor=blue,citecolor=Green,urlcolor=cyan}

\shorttitle{Barium Enrichment in Gaia MS+WD Binaries}
\shortauthors{Rekhi et al.}

\graphicspath{{./}{figures/}}

\begin{document}

\title{Ba Enrichment in Gaia MS+WD Binaries: Tracing \textit{s}-Process Element Production}

\correspondingauthor{Param Rekhi}
\email{param.rekhi@weizmann.ac.il}

\author[0009-0001-3501-7852]{Param Rekhi}
\affiliation{Department of Particle Physics and Astrophysics, Weizmann Institute of Science, Rehovot 7610001, Israel}

\author[0000-0001-6760-3074]{Sagi Ben-Ami}
\affiliation{Department of Particle Physics and Astrophysics, Weizmann Institute of Science, Rehovot 7610001, Israel}

\author[0000-0002-0430-7793]{Na'ama Hallakoun}
\affiliation{Department of Particle Physics and Astrophysics, Weizmann Institute of Science, Rehovot 7610001, Israel}

\author[0000-0001-9298-8068]{Sahar Shahaf}
\affiliation{Department of Particle Physics and Astrophysics, Weizmann Institute of Science, Rehovot 7610001, Israel}

\author[0000-0002-2998-7940]{Silvia Toonen}
\affiliation{Anton Pannekoek Institute for Astronomy, University of Amsterdam, 1090 GE Amsterdam, The Netherlands}

\author[0000-0003-4996-9069]{Hans-Walter Rix}
\affiliation{Max-Planck Institute for Astronomy, K\"onigstuhl 17, D-69117 Heidelberg, Germany}

\submitjournal{The Astrophysical Journal Letters}

\begin{abstract}
A large population of intermediate-separation binaries, consisting of a main-sequence (MS) star and a white dwarf (WD), recently emerged from Gaia's third data release (DR3), posing challenges to current models of binary evolution. Here we examine the \textit{s}-process element abundances in these systems using data from GALAH DR3. Following refined sample analysis with parameter estimates based on GALAH spectra, we find a distinct domain where enhanced \textit{s}-process elemental abundances depend on both the WD mass and metallicity, consistent with parameter spaces identified in previous asymptotic giant branch (AGB) nucleosynthesis studies having higher \textit{s}-process yields. Notably, these enhanced abundances show no correlation with the systems' orbital parameters, supporting a history of accretion in intermediate-separation MS+WD systems. Consequently, our results form a direct observational evidence of a connection between AGB masses and \textit{s}-process yields. We conclude by showing that the GALAH DR3 survey includes numerous Ba dwarf stars, within and beyond the mass range covered in our current sample, which can further elucidate \textit{s}-process element distributions in MS+WD binaries.
\end{abstract}


\section{Introduction}\label{sec:intro}

The study of binary systems consisting of a main sequence (MS) star and a white dwarf (WD) is fundamental to our understanding of stellar evolution and the complex processes governing mass transfer in binaries. Depending on the orbital separation in these systems, mass transfer can occur either through Roche-lobe overflow or common-envelope evolution, significantly impacting the evolutionary pathways of both stars involved \citep{Cehula_2023_TheoryMass}.
A detached  binary system may undergo a mass transfer phase as one of its components evolves. As it leaves the MS, the expanding star may fill its Roche lobe and lose some of its mass. The amount of mass loss and the efficiency with which the companion accretes it depends on the orbital parameters and stellar masses. Several possible mechanisms can govern mass transfer and influence the evolution of binary systems, from conservative funneling through the first Lagrange point to engulfment and inspiral \citep{Marchant_2024_EvolutionMassive}. 
While the specific details of binary evolution with mass transfer differ, several common traits emerge. First, the orbits of the resulting post-mass-transfer systems are expected to be circular. Second, energy and angular momentum exchange, either between the binary components or the interstellar medium, is expected to affect the distribution of orbital separations; specifically, regardless of the dominant mechanism, orbital separations of $\sim1$\,AU are expected to be disfavoured in post-mass-transfer binaries 
\citep{Zorotovic_2010_PostcommonenvelopeBinaries, Toonen_2012_SupernovaType, Toonen_2013_EffectCommonenvelope, Camacho_2014_MonteCarlo, Scherbak_2023_WhiteDwarf}.

Recent advancements in observational astronomy have been significantly propelled by data from the Gaia space observatory. The non-single star (NSS) \citep{Arenou_2023_GaiaData} catalogue in Gaia's third data release (DR3) provides detailed information on over 800,000 binary and multiple star systems. It includes astrometric, spectroscopic, and eclipsing binary data, offering insights into the properties and dynamics of these systems.
This wealth of astrometric binaries recently discovered by the Gaia mission provided an unexpected puzzle: thousands of MS+WD binaries separated by $\sim 1$\,AU, often exhibiting non-negligible eccentricities \citep{Shahaf_2023_TriageGaia,Shahaf_2024_TriageGaia}. Notably, available binary population synthesis (BPS) codes \citep{PortegiesZwart_1996_PopulationSynthesis, Nelemans_2001_PopulationSynthesis, Toonen_2012_SupernovaType} do not reproduce such a population \citep[see figure 16 of][]{Shahaf_2024_TriageGaia}. Since the progenitor asymptotic-giant-branch (AGB) stars of the WD companions also attain sizes of $\sim 1$\,AU, these binaries were presumably on the cusp of a common-envelope phase when the WD progenitor evolved. Therefore, some mass transfer processes probably occurred during their evolution. 
It is interesting to note that intermediate separation post-AGB+MS binaries with similar orbital parameters have previously been reported in literature \citep{Gorlova_2012_TimeresolvedSpectroscopy,Gorlova_2015_IRAS19135,Oomen_2018_OrbitalProperties}. \citet{Soker_2016_IntermediateLuminosity, Kashi_2018_CounteractingTidal} have proposed post-AGB binary interaction models resulting in such systems, which may be a step forward towards models for intermediate separation MS+WD binaries.

Barium stars were first described by \citet{Bidelman_1951_BaII} as G- and K-type giants with abnormally strong absorption lines of slow-neutron-capture (\textit{s})-process elements in their spectra. Although these stars exhibit enhanced abundances of quite a few \textit{s}-process elements, they have been historically referred to as Ba stars due to the remarkable strength of their Ba\,II lines \citep{Bidelman_1951_BaII, McClure_1984_BariumStars, Escorza_2019_BariumRelated}. Discovered a few decades later, the eponymous barium dwarfs are MS analogues of giant Ba stars, sharing the same chemical peculiarities \citep{North_1991_NatureSTR, North_1994_NatureSTR, Lu_1991_TaxonomyBarium, Lu_1983_CatalogSpectral, Escorza_2019_BariumRelated}.

The connection between binary accretion and the presence of \textit{s}-process elements in stellar photospheres is well-established \citep{McClure_1980_BinaryNature, McClure_1990_BinaryNature, Merle_2016_BaNot, Escorza_2019_BariumRelated, Roriz_2021_HeavyElements}. In brief, \textit{s}-process elements are synthesized and dredged up to the surfaces of thermally pulsing (TP)-AGB stars during shell burning \citep{Busso_1999_NucleosynthesisAsymptotic}. Accretion from the AGB star onto its companion transfers some of these \textit{s}-process elements. As surface \textit{s}-process abundances of MS (as well as RGB) stars closely track their metallicity \citep{Ahumada_2020_16thData, Buder_2021_GALAHSurvey}, an overabundance of these elements indicates past accretion from an AGB star, which later becomes a WD.

In this letter, we study a subsample of astrometric MS+WD binaries from Gaia \citep{Shahaf_2024_TriageGaia} with measured abundances of \textit{s}-process elements from the Galactic Archaeology with HERMES (GALAH) survey \citep{Ahumada_2020_16thData, Buder_2021_GALAHSurvey}. The sample is presented in Section~\ref{sec:sample}. In Section~\ref{sec:results} we discuss correlations between the Ba abundances and physical properties of these systems. Finally, in Section~\ref{sec:summary} we summarize and provide an outlook for ongoing and future projects.

\section{The Sample}\label{sec:sample}

\subsection{Gaia Astrometric Binaries}\label{sec:AMRF}

The third data release (DR3) of Gaia \citep{GaiaCollaboration_2023_GaiaData} included, for the first time, a catalog of unresolved sources with solutions for non-single stars \citep[NSS;][]{Arenou_2023_GaiaData}. \citet{Shahaf_2019_TriageAstrometric} have shown that in case the mass of the photometric primary star is known (e.g., based on its location on the Hertzsprung-Russell (HR) diagram), the nature of the companion can be further constrained, using the `Astrometric Mass-Ratio Function' (AMRF)---a unitless observational parameter defined as
\begin{equation}
    \mathcal{A} \equiv \frac{\alpha_0}{\varpi} \bigg(\frac{M_1}{\textrm{M}_\odot}\bigg)^{-1/3} \bigg(\frac{P}{\textrm{yr}}\bigg)^{-2/3}\, , 
    \label{eq:AMRF}
\end{equation}
where $\alpha_0$ is the angular semi-major axis of the photocenter, $\varpi$ is the parallax, $P$ is the orbital period, and $M_1$ is the mass of the photometric primary. The AMRF value derived from the observed parameters is then compared to the theoretical value for a valid set of mass ratios, $q=M_2/M_1$, and luminosity ratio, $\mathcal{S}=F_2/F_1$, assuming different companion types:
\begin{equation}\label{eq:expectedAMRF}
    \mathcal{A} = \frac{q}{(1+q)^{2/3}} \left( 1 - \frac{\mathcal{S}(1+q)}{q(1+\mathcal{S})} \right).
\end{equation}
Since luminous companions ($\mathcal{S}>0$) have a maximal possible AMRF value, if the observed AMRF is larger than this value, the companion is with a high probability a non-luminous object---likely a compact one \citep[see][]{Shahaf_2023_TriageGaia}. Similarly, a boundary curve can be drawn between the expected AMRF values of MS+MS binaries and hierarchical triple MS systems for a given primary mass: If the primary is an MS star, systems where the companion is another MS+MS binary can get to larger AMRF values compared to systems with single MS companions \citep{Shahaf_2019_TriageAstrometric}. In this range of AMRF values a single MS companion is ruled out, and the companion can be either an inner MS+MS binary or a compact object. \citet{Shahaf_2024_TriageGaia} have shown how Gaia synthetic photometry (calculated based on Gaia's low-resolution BP/RP spectra) can be used to further differentiate between systems with red excess flux, where the companion is likely an MS+MS binary, and systems with no red excess flux, where the companion is likely a compact object. Given the measured dynamical masses of the companions in these systems, the companions are most likely WDs. For WD secondaries, $\mathcal{S}$ approaches 0. In such a case, $q$, and hence \m2, can be uniquely determined from the AMRF as the solution of the equation 
\begin{equation}
    \mathcal{A} = \frac{q}{(1+q)^{2/3}}.
    \label{eq:AMRF M2}
\end{equation}

Using the AMRF triage method, \citet{Shahaf_2024_TriageGaia} identified a population of $\approx 9,800$ astrometric binaries from the Gaia where the primary is a $<1.2$\,\msun\ MS star, and the secondary is unlikely to be a single MS star. Based on Gaia synthetic photometry they concluded that nearly 3,200 of these are probable MS+WD binaries (i.e., show no red excess flux).

\subsection{GALAH}

GALAH is a high-resolution ($R\approx 28,000$) spectroscopic survey conducted with the HERMES \citep{Sheinis_2015_FirstLight} spectrograph coupled to the 3.9\,m Anglo-Australian Telescope in New South Wales, Australia. The survey conducted observations between 2013 and 2019 in four narrow bands ($\sim 200$\,\AA) ranging from blue to near-infrared. GALAH DR3 provides one-dimensional spectra, stellar atmospheric parameters and elemental abundances for nearly 600,000 stars within a few kpc of the Sun \citep{Buder_2021_GALAHSurvey}.
For each star it lists abundances for multiple \textit{s}-process elements including Ba, Y, Zr and La. Spectra for all stars listed in the catalogue are publicly available and can be downloaded from the GALAH DR3 website\footnote{\url{https://www.galah-survey.org/dr3/overview}}. 

To build our sample, we identify Gaia DR3 sources common to the AMRF MS+WD sample of \citet{Shahaf_2024_TriageGaia} and the GALAH DR3 catalogue. To ensure the reliability of the GALAH parameters we use, we implement quality cuts on the GALAH catalogue as follows: We keep GALAH sources having stellar parameter, iron abundance and alpha abundance quality flags (\code{flag\_sp}, \code{flag\_fe\_h} and \code{flag\_alpha\_fe}) equal to 0. We also keep sources with \code{flag\_sp} value of 1, as this flag indicates high Gaia RUWE\footnote{Renormalized Unit Weight Error \citep[see][]{GaiaCollaboration_2023_GaiaData}} values, which are on par for our sample given that it consists of astrometric binaries. Detailed explanations of all GALAH quality flags as well as recommended quality cuts are given on their website\footnote{\url{https://www.galah-survey.org/dr3/flags/}}\textsuperscript{,}\footnote{\url{https://www.galah-survey.org/dr3/using_the_data/}\label{f:galahrec}}.
A small fraction (15 objects) of the cross-matched stars are identified by GALAH as possible binary systems (\code{flag\_sp} $~= 64$) based on their aberrant positions on stellar evolutionary tracks. While this is also not surprising for our sample, we exclude these stars as their spectral parameters are known by GALAH to be error prone. These stars are also redder than the MS track, indicating they are likely hierarchical triples. Furthermore, only one of these stars shows enhanced Ba abundance and ten of them were classified as having red color excess by \citet{Shahaf_2024_TriageGaia}.

\subsubsection{Control sample}\label{sec:control}

We define a control sample of $\approx 1.8 \times 10^5$ MS field stars selected from GALAH DR3. The control stars are chosen such that they follow the quality cuts recommended by the GALAH DR3 survey\textsuperscript{\ref{f:galahrec}}, and occupy the same $T_\text{eff} - \log g$ space as the MS stars in the AMRF-GALAH sample.

\begin{figure}
    \centering
    \subfloat[]{
        \includegraphics[width=\columnwidth]{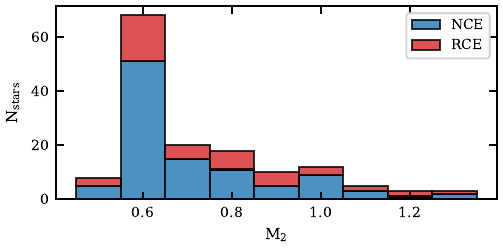}
}\\
    \subfloat[]{
    \includegraphics[width=\columnwidth]{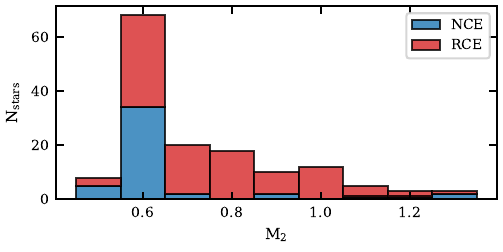}
    }
    \caption{Stacked histograms showing the distribution of GALAH derived \m2 in the NCE and RCE samples derived from (a) GALAH DR3 parameters (this work) and (b) Gaia DR3 parameters \citep{Shahaf_2024_TriageGaia}. More than 50\% of stars with \m2 $\approx 0.6$ \msun formerly classified as RCE get reclassified as NCE systems.}
    \label{fig:m2 comp}
\end{figure}

\subsection{Recalculating the AMRF, \m2 and Color Excess Values}\label{sec:reAMRF}

As given in Equation~(\ref{eq:AMRF}), the calculation of the AMRF, and subsequently the secondary mass (\m2), requires knowledge of the mass of the photometric primary (\m1). The red excess flux can in turn be determined based on the location of the system on the HR diagram compared to a theoretical isochrone of a single MS star of a given age, metallicity, and extinction.
\citet{Shahaf_2024_TriageGaia} relied on catalogs and maps published based on Gaia data for the values of the primary mass, the metallicity and the extinction. The primary masses were estimated based on the location on the HR diagram by \citet{Arenou_2023_GaiaData}, and the metallicities were estimated using Gaia's low-resolution BP/RP spectra by \citet{Zhang_2023_Parameters220}. The age was taken as a fixed value of 2\,Gyr, since the location on the HR diagram is not expected to change significantly throughout the MS lifetime, and based on indications for a star formation burst having happened $\sim 2$\,Gyr ago \citep{Mor_2019_GaiaDR2}.

Here, we utilize stellar parameters provided by GALAH to recalculate the AMRF, corresponding \m2 and color excess values.
As Gaia low-resolution BP/RP spectra cannot isolate most spectral lines, and Gaia masses do not take into account stellar metallicity, we prefer GALAH measurements of [Fe/H] and \m1 to those obtained with Gaia, further utilizing GALAH age estimates as outlined below.

We analytically obtain updated values of the AMRF and \m2 using Equations (\ref{eq:AMRF}) and (\ref{eq:AMRF M2}) and primary masses derived by GALAH. 
We then utilize GALAH's age and metallicity estimates to recalculate the color excess: We use Gaia synthetic photometry \citep{Gaia_2023_SynPhot} and PARSEC\footnote{PARSEC v1.2S, available online via \url{http://stev.oapd.inaf.it/cmd}.} \citep{Bressan_2012_PARSECStellar, Tang_2014_NewPARSEC, Chen_2015_ParsecEvolutionary} isochrones to compute the Johnson-Kron-Cousins $B-I$ color excess,
\begin{equation}
\Delta \left(B-I\right) = \left(B-I\right)_\text{observed}-\left(B-I\right)_\text{expected},
\label{eqn: rce}
\end{equation}
where $(B-I)_\text{observed}$ is the observed dereddened $B-I$ colour index, and $(B-I)_\text{expected}$ is the expected colour index obtained from the PARSEC isochrones.
We use PARSEC isochrones to create a 3-D linear interpolant model\footnote{We use the \noun{scipy} class \code{LinearNDInterpolator}, documented at \url{https://docs.scipy.org/doc/scipy/tutorial/interpolate.html}.} for the expected color excess as a function of stellar absolute $V$-band magnitude, metallicity and log(age). We compute the GALAH metallicity estimate from the iron and $\alpha$ abundances using the formula 
\begin{eqnarray}
\log \left(\frac{Z}{Z_{\odot }}\right) & = & {\rm [Fe/H]} + \log (10^{[\alpha /{\rm Fe}]} 0.694+ 0.306)
\end{eqnarray}
given by \citet{Salaris_2005_EvolutionStars}.
To ensure the interpolant is smooth and representative of our sample, we mask the portions of the isochrones representing stellar evolution beyond the MS.
We exclude four stars from the sample which have age estimates lower than the expected MS progenitor lifetimes corresponding to their respective secondary (WD) mass. Following \citet{Shahaf_2024_TriageGaia}, we account for interstellar extinction using the dust maps of \citet{Green_2019_3DDust} and \citet{Lallement_2019_Gaia2MASS3D}. The $E(B-V)$ values given in the maps are converted to $B$ and $I$ bands using extinction coefficients $R_B$, $R_I$, and $R_V$ of 3.626, 1.505, and 3.1, respectively \citep{Schlafly_2011_MeasuringReddening}.

To account for the effect of uncertainties in the input parameters on the color excess, we run a Monte Carlo procedure with $10^5$ realizations for each star, with the final color excess being the median of all successful realizations. The input parameters $V$ mag, [M/H] and log(age) are randomly drawn from uncorrelated normal distributions having the observed parameter means and standard deviations.For some stars, a fraction of realizations are not successful due to one or more input parameters falling outside the interpolant boundaries. This forms the basis of the final quality cut to our sample wherein we exclude stars which have less than 90\% successful realizations. 
\begin{deluxetable}{lcc}
\label{tab:sample}
\tablecaption{Number of MS+WD systems in GALAH DR3. Numbers of stars with enhanced Ba abundance are indicated. The final sample used in this letter is the one with red color excess $<0.01$. See Section~\ref{sec:sample} for details of the various categories.}
\tablehead{\colhead{} & \colhead{All} & \colhead{\bafe\ $>0.25$}} 
\startdata
Total & 216 & 54 \\
Post quality cut & 166 & 44 \\
Post interpolation & 147 & 39 \\
Red color excess $<0.01$ & 102 & 26\\
\enddata
\end{deluxetable}

MS+WD systems are classified as having a red color excess smaller than 0.01 mag, which forms the minimal color excess for hierarchical triples (see figure~2 of \citealt{Shahaf_2024_TriageGaia}).
We find 102 such systems in the AMRF-GALAH sample, a significantly different number from the previous estimate of 47 systems in the same sample obtained by \citet{Shahaf_2024_TriageGaia}.
Furthermore, a bulk of the stars having \m2 near the canonical WD mass of $\approx 0.6$ \msun\ and formerly classified as having red-color-excess (RCE) get reclassified as no-red-color-excess (NCE) systems, see Figure~\ref{fig:m2 comp}. We attribute this to the metallicity and age estimates enabled by GALAH spectra.
We note, however, that some MS+WD systems might still be classified as having red color excess, possibly due to still inaccurate mass, age, metallicity and extinction estimates.
Table~\ref{tab:sample} lists the number of systems in the initial cross-match sample as well as the cuts leading to the final NCE sample, which consists of 102 systems. We exclusively use the NCE sample in the following sections.

\section{Results and Discussion}\label{sec:results}

In this section, we explore the relation between the \textit{s}-process abundances of the MS primaries and the stellar and orbital parameters of the system.
\begin{figure}
    \centering
    \includegraphics[width=\columnwidth]{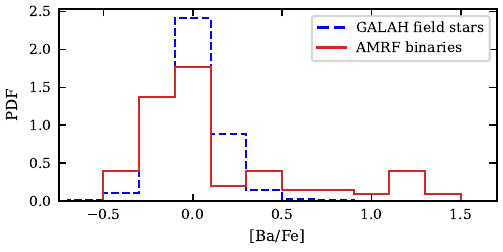}
    \\
    \includegraphics[width=\columnwidth]{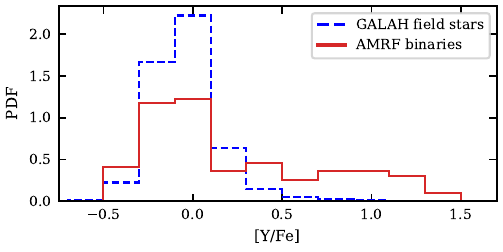}
    \caption{Distributions of \bafe\ and \yfe\ abundances in the AMRF-GALAH sample. The abundance distribution of field stars in GALAH DR3 is shown in dashed-blue. A K-S test shows that both the AMRF Ba and Y abundances are inconsistent with the field star distributions with $p$-values of less than $10^{-6}$.}
    \label{fig:ba y pdf}
\end{figure}
We consider enhanced Ba and Y abundances as representative of \textit{s}-process enrichment, as other \textit{s}-process elements (Sr, Zr, La, Ce, Rb etc) are not measured reliably by GALAH for a large fraction of our sample\footnote{\url{https://www.galah-survey.org/dr3/caveats/\#high-abundances-of-v-co-rb-sr-zr-mo-ru-la-nd-and-sm}}. This is justified by the strong correlation between all \textit{s}-process element abundances seen in enhanced systems where measurements do exist, see Figure~\ref{fig: s-proc corner}. We only consider abundance measurements considered reliable by GALAH, i.e, having a quality flag (\code{flag\_X\_fe} for element X) equal to 0. All 102 stars in our sample have reliable Ba abundance measurements, out of which 98 also have reliable Y measurements.

Figure~\ref{fig:ba y pdf} shows the \bafe\ and \yfe\ abundance distribution in the AMRF-GALAH sample compared to that of the control sample (see Section~\ref{sec:control}).
The AMRF-GALAH sample has a significant tail favoring enhanced \textit{s}-process abundances ($\text{\bafe, \yfe} \gtrsim 0.25$), clearly diverging from the field star distribution. A Kolmogorov-Smirnov (K-S) test shows that both the Ba and Y abundances of the AMRF-GALAH sample are inconsistent with the field star distributions with $p$-values smaller than $10^{-6}$.

\begin{figure}
    \centering
    \subfloat[]{
        \includegraphics[width=\columnwidth]{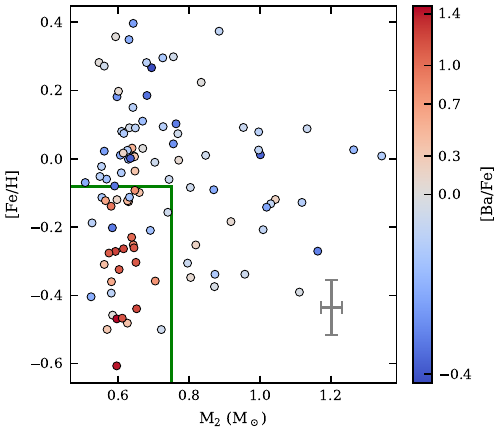}
        \label{subfig: feh_m2_BaFe}}\\
    \subfloat[]{
        \includegraphics[width=\columnwidth]{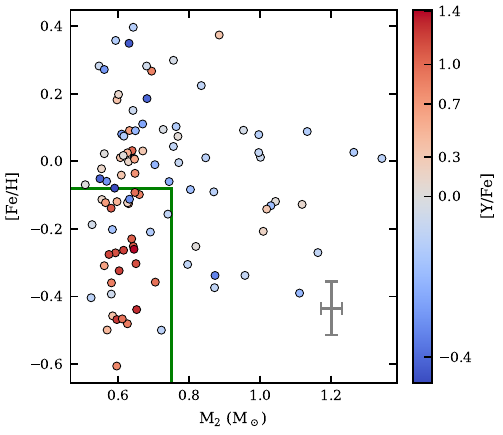}
        \label{subfig: feh_m2_YFe}}
    \caption{(a) Barium and (b) yttrium abundance as a function of iron abundance of the primary and mass of the secondary (which we assume is a WD). Median error bars are shown in grey. We note that enhanced Ba abundances are found almost exclusively in the region of \m2 $<0.75$ and [Fe/H] $<-0.08$ (bounded in green), with Y enhanced stars extending to higher metallicities. }
    \label{fig: feh_m2}
\end{figure}

\begin{figure}
    \centering
    \includegraphics[width=\columnwidth]{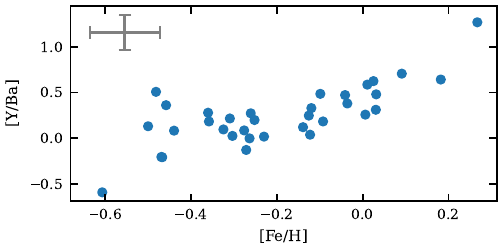}
    \caption{Comparison of Ba and Y abundances as a function of iron abundance. Stars are limited to the region of \m2 $<0.75$ and [Y/Fe] $>0.25$.
    Y, being a light (first-peak) s-process element is predicted to be produced more efficiently at higher [Fe/H] values as compared to the heavier (second-peak) Ba \citep{Busso_2001_NucleosynthesisMixing}. We find [Y/Ba] to increase with [Fe/H], agreeing with the prediction.}
    \label{fig:Ba Y comp}
\end{figure}

We find that the Ba-enhanced stars ([Ba/Fe]~$>0.25$) are predominantly located in the region where \m2~$<0.75$ \msun\ and [Fe/H]~$<-0.08$ (Figure~\ref{fig: feh_m2}). Y-enhanced stars follow the same general trend, albeit with a heavy tail extending into higher metallicities. 
This follows previous studies showing the dependence of \textit{s}-process production in the AGB phase on the progenitor mass and metallicity. Nucleosynthesis models by \citet{Karakas_2016_StellarYields} show that Ba yields in the AGB phase peak at a progenitor mass of $\approx 2$\,\msun\ (WD mass $\approx 0.65$\,\msun), plateauing till $\approx 4$\,\msun (WD mass $\approx 0.9$\,\msun), and falling substantially for higher masses. This in turn implies that Ba abundances in the progenitor steadily decrease for WDs more massive than $\approx 0.65$\,\msun. Furthermore, recent observational studies by \citet{Roriz_2021_HeavyElements} and \citet{Vilagos_2024_BariumStars} demonstrate an anti-correlation of AGB \textit{s}-process abundances with metallicity, with a considerable suppression for super-solar metallicities. The presence of Y enhanced stars at higher metallicities ([Fe/H] $\lesssim 0.2$; Figure~\ref{subfig: feh_m2_YFe}) also conforms with the the aforementioned studies: Y is a first-peak (lighter) \textit{s}-process element, unlike Ba which belongs to the second peak, and has a weaker dependence on metallicity \citep[Figure \ref{fig:Ba Y comp};][]{Busso_2001_NucleosynthesisMixing}.
Our results thus seem to indicate that we observe enhanced \textit{s}-process abundances in the AMRF photometric primaries for almost all cases where substantial amounts of \textit{s}-process elements were generated in the WD progenitor. This in turn implies that, even though they do not show \textit{s}-process enrichment, the higher metallicity MS+WD systems in our sample could also have undergone accretion in the past.

\begin{figure*}[p]
    \centering
    \subfloat[]{
        \includegraphics[width=\columnwidth]{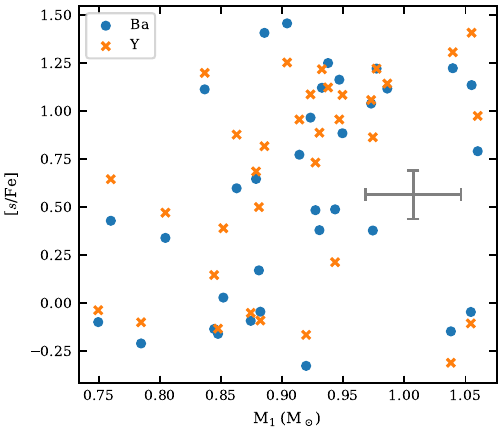}
        }
    \subfloat[]{
        \includegraphics[width=\columnwidth]{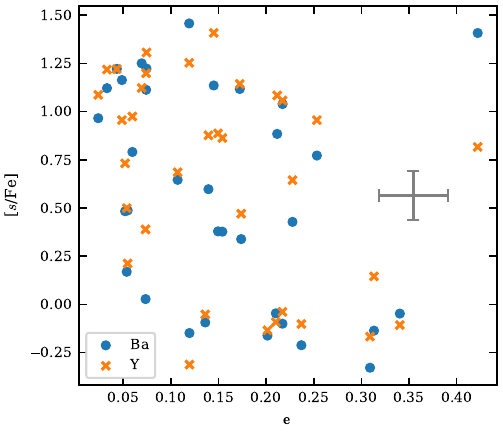}
        }
    \\
    \subfloat[]{
        \includegraphics[width=\columnwidth]{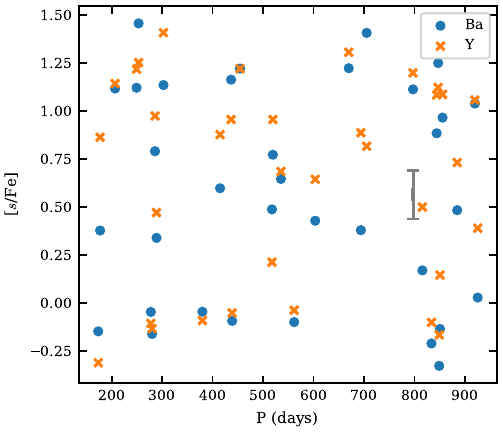}
        }
    \subfloat[]{
    \includegraphics[width=\columnwidth]{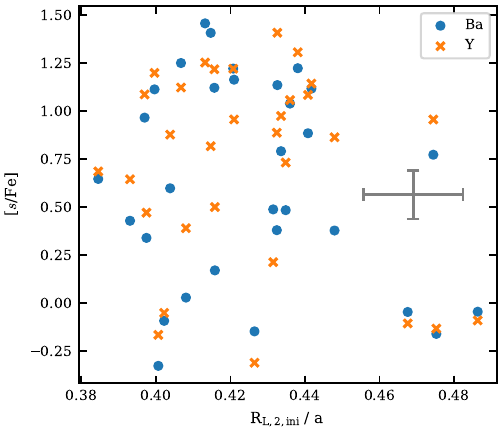}
    }
    \caption{Barium and yttrium abundances as a function of (a) primary mass, (b) eccentricity, (c) period and (d) Roche-lobe radius of the WD progenitor normalized by the orbital separation. Median error bars are shown in grey. The plotted systems belong to the region \m2~$<0.75$ and [Fe/H]~$<-0.08$ (see Figure~\ref{fig: feh_m2} and text for explanation). Although Y-enriched stars extend to high metallicities, we define the \textit{s}-process domain as per Ba abundances following existing convention.
    Roche-lobe radii are computed using the formula given by \citet{Eggleton_1983_AproximationsRadii}, with the initial-final mass relation used to obtain progenitor masses taken from \citet{Cummings_2018_WhiteDwarf}.}
    \label{fig: Ba_fe orbital params}
\end{figure*}

Further supporting this view is our observation of no substantial correlation of Ba and Y abundances with the orbital parameters of the system. Figure~\ref{fig: Ba_fe orbital params} shows the Ba abundance as a function of the primary mass, eccentricity, period and Roche lobe radius of the WD progenitor, restricted to the \textit{s}-process enhanced region determined previously.
Due to their origins from Gaia astrometry, the maximal orbital period of these binaries is $\mathcal{O}(10^3)$ days, which constrains their separations to $\mathcal{O}(1\,\text{AU})$ \citep{GaiaCollaboration_2023_GaiaData, Arenou_2023_GaiaData}. Previous studies of Ba giants have shown some period relations only post $10^3$ days, with more or less flat distribution for shorter periods \citep[for eg., see figure 13 of ][]{Escorza_2019_BariumRelated}. As shown in Figure \ref{fig: Ba_fe orbital params}, our sample also exhibits a flat period distribution of enhanced \textit{s}-process abundances. 
We conclude that at the intermediate separations probed by our sample, binary accretion occurs with a high probability, and the principal parameters that determine the \textit{s}-process enhancement of the photometric primary are inherent to the secondary progenitor (its mass and metallicity).
Furthermore, our results give evidence for the relative decrease in Ba production as a function of stellar mass in the AGB phase.

We find it important to insert a cautionary note here: the \textit{s}-process enhanced domain presented in this work is purely empirical and, due to the small sample size, is indicative rather than conclusive. Furthermore, the AMRF sample from \citet{Shahaf_2024_TriageGaia} is biased toward higher eccentricities with increasing distances (see figure 12 of \citealt{Shahaf_2024_TriageGaia}), which corresponds to a similar bias related to primary/secondary mass (as massive primaries are more easily characterized at larger distances). This effect is compounded by a high-mass deficit in the sample \citep{Hallakoun_2024_DeficitMassive}. This bias, and its corollary—lower eccentricities for lower masses—is also reflected in the AMRF-GALAH cross-matched sample. Consequently, we are unable to test the independence of the \textit{s}-process enrichment to orbital parameters at the extremes of their parameter space.

\section{Summary and Outlook}\label{sec:summary}

\begin{figure}
    \centering
    \includegraphics[width=\columnwidth]{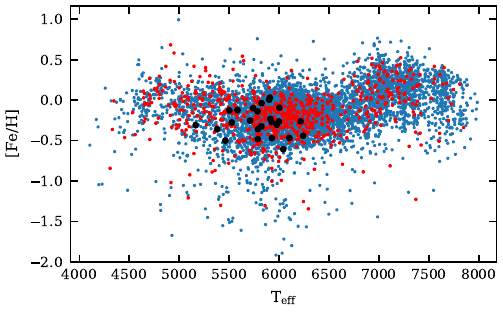}
    \\
    \includegraphics[width=\columnwidth]{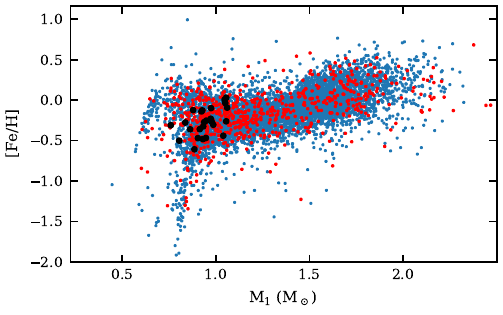}
    \caption{Barium and yttrium enriched dwarf stars ($\text{\bafe, \yfe} > 0.25$) from the GALAH DR3 survey. The blue and red points have Gaia RUWE values less than and greater than 1.4, respectively. The black dots are the enriched stars in the AMRF-GALAH NCE sample.}
    \label{fig:galah ba dwarfs}
\end{figure}

In this paper, we leverage the Gaia NSS and GALAH DR3 catalogues to study the \textit{s}-process abundances of intermediate-separation MS+WD systems. Following the procedure laid out by \citet{Shahaf_2024_TriageGaia}, we filter the AMRF-GALAH sample to remove contamination from hierarchical triples based on their $B-I$ color excess. We find the metallicity and age estimates derived from high-resolution GALAH spectra to significantly increase the accuracy of this classification.

Our results reveal a distinct domain where enhanced \textit{s}-process elements depend on both WD mass and metallicity. This same domain has been identified in previous AGB nucleosynthesis studies as producing the highest \textit{s}-process yields. Additionally, the enhanced abundances show no correlation with the orbital parameters of the systems. Consequently, our results are consistent with a history of accretion in intermediate-separation MS+WD systems. Furthermore, we offer direct evidence linking AGB masses to \textit{s}-process yields, enhancing our understanding of the underlying stellar processes.

Looking forward, our sample sets the stage for further advancements in binary evolution and AGB yield simulations. However, to refine these models, larger and diverse samples are necessary.
We are actively pursuing additional spectroscopic follow-ups of the AMRF MS+WD sample using FEROS \citep{Kaufer_1998_FEROSNew,Kaufer_1999_CommissioningFEROS}. 
Furthermore, as shown in Figure~\ref{fig:galah ba dwarfs}, the GALAH DR3 survey contains a wealth of Ba and Y enriched dwarf stars within the same phase space as covered in this work, also extending into higher masses and temperatures not covered by the current sample of \citet{Shahaf_2024_TriageGaia}.
We note that while Ba is a classical indicator of \textit{s}-process enhancement, studies have shown that it may appear to be anomalously enhanced compared to other s-process elements due to non-nucleosynthetic phenomena \citep{Spina_2020_HowMagnetic, Baratella_2021_GaiaESOSurvey}. We hence recommend corroboration of Ba abundances with other s-process elements as demonstrated here. We plan to explore this sample of \textit{s}-process enriched main-sequence stars in future studies, which promises to further elucidate the dependencies and distributions of \textit{s}-process elements in MS+WD binaries.

\begin{acknowledgments}
S.B.A. is grateful for support from the Azrieli Foundation, André Deloro Institute for Advanced Research in Space and Optics, Peter and Patricia Gruber Award, Willner Family Leadership Institute for the Weizmann Institute of Science, Aryeh and Ido Dissentshik Career Development Chair, Israel Science Foundation (grant 1920/21), Israel Ministry of Science (grant 3-18140), and Minerva Stiftung. 
The research of S.S. is supported by a Benoziyo prize postdoctoral fellowship.
S.T. acknowledges support from the Netherlands Research Council NWO (VIDI 203.061 grant).

This work made use of the Third Data Release of the GALAH Survey \citep{Buder_2021_GALAHSurvey}. The GALAH Survey is based on data acquired through the Australian Astronomical Observatory, under programs: A/2013B/13 (The GALAH pilot survey); A/2014A/25, A/2015A/19, A2017A/18 (The GALAH survey phase 1); A2018A/18 (Open clusters with HERMES); A2019A/1 (Hierarchical star formation in Ori OB1); A2019A/15 (The GALAH survey phase 2); A/2015B/19, A/2016A/22, A/2016B/10, A/2017B/16, A/2018B/15 (The HERMES-TESS program); and A/2015A/3, A/2015B/1, A/2015B/19, A/2016A/22, A/2016B/12, A/2017A/14 (The HERMES K2-follow-up program). We acknowledge the traditional owners of the land on which the AAT stands, the Gamilaraay people, and pay our respects to elders past and present. This paper includes data that has been provided by AAO Data Central (\url{datacentral.org.au}).

This work has made use of data from the European Space Agency (ESA) mission
{\it Gaia} (\url{https://www.cosmos.esa.int/gaia}), processed by the {\it Gaia} Data Processing and Analysis Consortium (DPAC, \url{https://www.cosmos.esa.int/web/gaia/dpac/consortium}). Funding for the DPAC has been provided by national institutions, in particular the institutions participating in the {\it Gaia} Multilateral Agreement.
\end{acknowledgments}

\software{Astropy \citep{AstropyCollaboration_2013_AstropyCommunity, AstropyCollaboration_2018_AstropyProject, AstropyCollaboration_2022_AstropyProject}, 
Matplotlib \citep{Hunter_2007_Matplotlib2D}, 
NumPy \citep{Harris_2020_ArrayProgramming}, 
Pandas \citep{McKinney_2010_DataStructures, ThePandasDevelopmentTeam_2024_PandasdevPandas},
SciPy \citep{Virtanen_2020_SciPyFundamental}, 
uncertainties \citep{EricO.Lebigot_2024_UncertaintiesPython}}

\facilities{\textit{Gaia}, AAT}

\clearpage
\appendix
\restartappendixnumbering

\onecolumngrid 
\section{Correlations between abundance measurements of \textit{s}-process elements} \label{sec:appendixA}

\begin{figure*}[h]
    \centering
    \includegraphics[height=0.78\textheight]{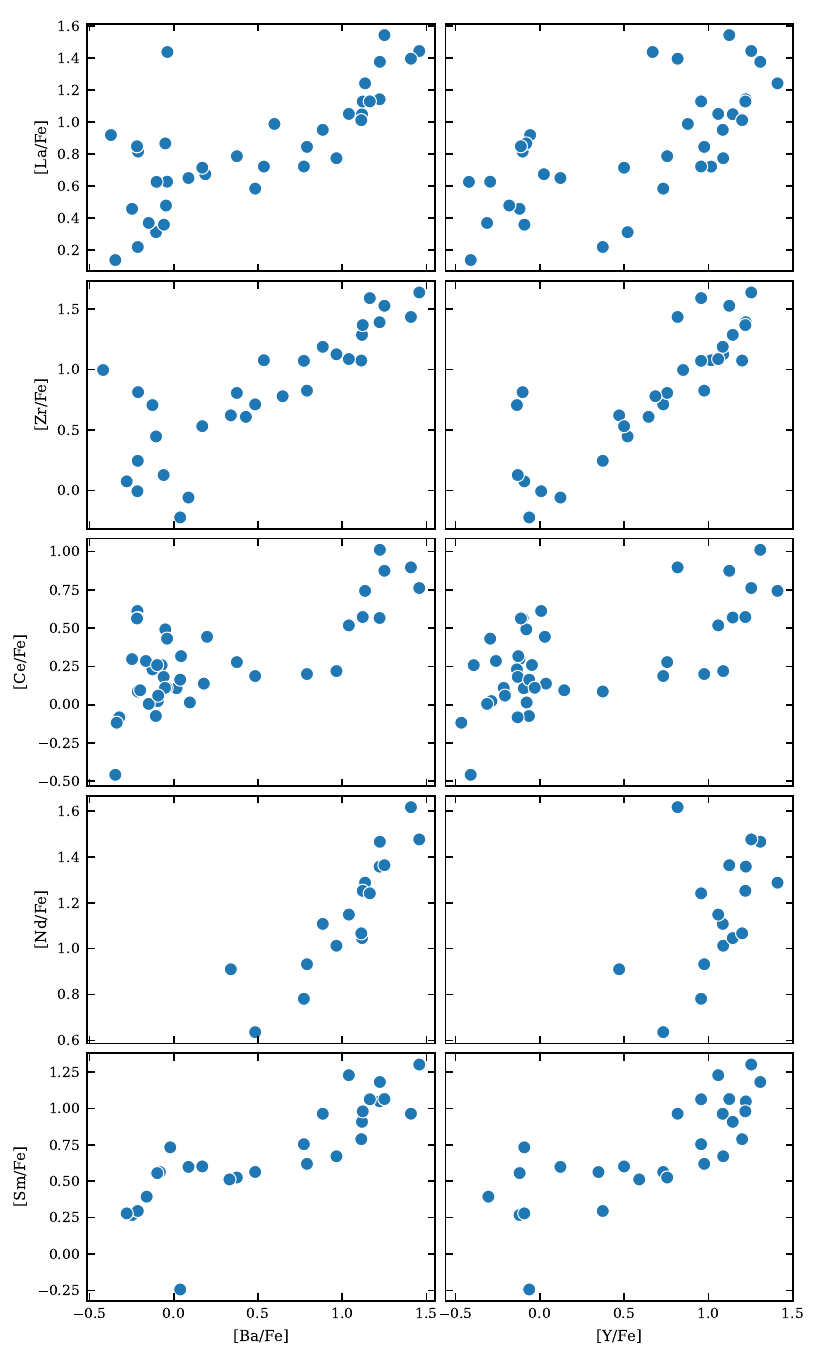}
    \caption{Correlations between abundance measurements of \textit{s}-process elements in the NCE sample. Although, GALAH flags high abundances of \textit{s}-process elements other than Ba and Y as possibly unreliable, we demonstrate here that their abundances are positively correlated with Ba and Y, justifying their use as representative of the \textit{s}-process in this work. }
    \label{fig: s-proc corner}    
\end{figure*}

 In Figure \ref{fig: s-proc corner}, we demonstrate strong correlations between enhanced \textit{s}-process element abundances in the NCE sample.

\twocolumngrid
\bibliography{Galah_AMRF_group}
\bibliographystyle{aasjournal}

\end{document}